\documentstyle[preprint,prl,tighten,aps,epsfig]{revtex}

\begin{document}

\draft

\preprint{\begin{tabular}{l}
\hbox to\hsize{\mbox{ }\hfill KIAS--P99102}\\
\hbox to\hsize{\mbox{ }\hfill hep--ph/9910557}\\
\hbox to\hsize{\mbox{ }\hfill November 1999}\\
          \end{tabular}}

\title{MSSM Higgs--Boson Production at Hadron Colliders \\
       with Explicit CP Violation}

\author{S.Y. Choi and Jae Sik Lee}
\address{Korea Institute for Advanced Study, 207--43, Cheongryangri--dong
         Dongdaemun--gu, Seoul 130--012, Korea}

\maketitle

\begin{abstract}
Gluon fusion is the main production mechanism for Higgs bosons with masses
up to several hundred GeV in $pp$ collisions at the CERN Large Hadron 
Collider. We investigate the effects of the $CP$--violating phases 
on the fusion process including both the sfermion--loop contributions 
and the one--loop induced $CP$--violating scalar--pseudoscalar mixing 
in the minimal supersymmetric standard model.  With a universal trilinear 
parameter assumed, every physical observable involves only the sum of 
the phases of the universal trilinear parameter $A$ and the higgsino 
mass parameter $\mu$. The phase affects the lightest Higgs--boson 
production rate significantly through
the neutral Higgs--boson mixing and, for the masses around the 
lightest stop--pair threshold, it also changes the production
rate of the heavy Higgs bosons significantly  through both the stop 
and sbottom loops and the neutral Higgs--boson mixing. 
\end{abstract}

\pacs{PACS number(s): 11.30.Er, 12.60.Jv, 13.10.+q}


\section{Introduction}
\label{sec:introduction}

The experimental observation of Higgs bosons and the
detailed confirmation of their fundamental properties are crucial 
for our understanding of the mechanism responsible
for the electroweak symmetry breaking and they constitute one of the most 
important experimental programme at the CERN Large Hadron Collider (LHC), 
the physics potential of
which has been investigated with a considerable amount of effort \cite{LHC}.
Within the minimal supersymmetric standard model (MSSM), the
lightest Higgs boson is expected to be well within the reach of the 
LHC and also the heavier Higgs bosons with their masses less than 
1 TeV could be found at the LHC.  Gluon fusion \cite{GGMW}
is the main production mechanism for Higgs bosons with masses up to 
several hundred GeV  at the LHC. The dominant production of 
these three neutral Higgs bosons, which proceeds through loops of both 
quarks and squarks (primarily, those of top and bottom flavor),  
can be significantly affected by non--zero 
$CP$--violating phases in the MSSM \cite{MS,DeMo}, associated 
with the higgsino 
mass parameter $\mu$ and trilinear
scalar couplings $A_f$. 

It has recently been pointed out \cite{IN} that these $CP$--violating
phases do not have 
to be suppressed in order to satisfy the constraints from electron and 
neutron electric dipole moments (EDMs). In this context, the so--called 
effective supersymmetry model \cite{KAPLAN} is of particular interest, where
sfermions of the first and second generations are decoupled, but
sfermions of the third generation remain relatively light. 
Based on the scenarios of this type, many important works  
in $B$ decays \cite{Many1}, dark--matter searches, and 
collider experiments \cite{Many2,DEMIR,Many3} have been recently reported. 
In the present work, we concentrate on the 
effects of the $CP$--violating phases on the Higgs--boson production
in gluon--gluon fusion at the LHC including both the sfermion--loop 
contributions \cite{DeMo} and the neutral Higgs--boson mixing \cite{PW}.

In this scheme, the characteristic $CP$--violating phenomena in the Higgs
sector are a possibly large mixing between the $CP$--even and $CP$--odd
neutral Higgs bosons and an induced relative phase $\xi$ 
\cite{DEMIR,PW} between the vacuum expectation values of the two Higgs
doublets. In particular, the heavy $CP$--even and $CP$--odd Higgs bosons
experience a maximal mixing in most cases due to their sharp 
mass degeneracy, in particular for large masses. In this light, 
the neutral Higgs--boson mixing cannot be simply
neglected with the argument that it is induced at the one--loop level.
Neglecting the other small loop contributions from the chargino and 
neutralino sectors,  the Higgs potential of the MSSM has one--loop 
radiative corrections mainly from the stop and sbottom sectors due to the 
large $t$ and $b$ Yukawa couplings.

On the other hand, the contributions of the sfermion loops to the 
gluon--gluon--Higgs vertices are expected to be large for sfermion 
masses similar to or smaller than the Higgs--boson mass. 
Concerning these contributions, Dedes and Moretti \cite{DeMo} 
have shown that the squark loop content of the dominant Higgs production 
mechanism via the gluon--gluon fusion mode at the LHC could be responsible 
for large corrections to the known cross sections. However, 
the possible large modifications due to the neutral Higgs--boson
mixing are not discussed in their work. 

We re--visit the important production mechanism
of Higgs bosons via gluon--gluon fusion at hadron colliders including 
both the stop/sbottom loop diagrams and the $CP$--violating mixing 
among three neutral Higgs bosons, and provide a detailed analysis of
the effects of the $CP$--violating phases on the Higgs--boson 
production  rates. The one--loop Feynman diagrams contributing to
the process $gg\rightarrow H_i$ in the MSSM are shown in Fig.~1.

{}From the viewpoint of radiative corrections, the neutral Higgs--boson 
mixing contributes to the Higgs--boson production in gluon--gluon 
annihilation at the two--loop level. For the sake of consistency,
it might be necessary to take into account all the two--loop vertex 
corrections to the gluon--gluon--Higgs interactions. 
However, the most dominant radiative corrections are expected to be
largely contained in the propagators and the vertex corrections are 
much smaller in size compared to the propagator corrections \cite{PR}. 
This point has been demonstrated explicitly in the Higgs--boson pair
production process $e^+e^-\rightarrow H_iH_j$ by Demir \cite{HH}.
Moreover, the sharp degeneracy of the heavy $CP$--odd and $CP$--even Higgs
states causes a large mixing between two heavy Higgs--boson states
even for a tiny $CP$--violating correction to the Higgs--boson mass matrix.
In this light, we could neglect the two--loop vertex corrections 
to the gluon--gluon--Higgs interactions to a good approximation
and concentrate on the effects of the neutral Higgs--boson mixing 
modifying the Higgs--boson propagators. Clearly, it will be important
to demonstrate quantitatively through explicit calculations that 
the two--loop vertex corrections are negligible compared to the
neutral Higgs--boson mixing in the Higgs--boson propagators, even if
we do not touch on the comparison in the present work.

The paper is organized as follows. In Sec.~II we describe
our theoretical frameworks; the MSSM Lagrangian, the neutral Higgs--boson
mixing and the sfermion mixing, including a brief discussion
of experimental limits on the parameters of the model. 
Sect.~III is devoted to the general description of the Higgs production 
via gluon--gluon fusion at hadron colliders. In Sec.~IV we present 
our numerical results.
And then we summarize our findings and draw conclusions in Sec.~V.
Finally, the Appendix is devoted to listing  all the Feynman rules 
relevant for our numerical analysis.

\section{CP Violation in the MSSM}

\subsection{The MSSM Lagrangian}

In this section, we list the relevant MSSM Lagrangian terms including
$CP$--violating sources, and the effective Lagrangian for the Higgs potential 
with radiative interactions to quartic couplings coming from enhanced
Yukawa couplings of the third generation. 
For a general purpose, it is useful to replace two conventional MSSM Higgs 
doublet fields $H_1$ and $H_2$ by $\Phi_1=+i\tau_2\,H^*_1$ and $\Phi_2$
following the notations used in Ref.~\cite{PW}. 
Then, the relevant MSSM Lagrangian is divided into four parts:
\begin{eqnarray}
-{\cal L}_{\rm soft}&=&\tilde{m}^2_Q\tilde{Q}^\dagger\tilde{Q}
                      +\tilde{m}^2_U\tilde{U}^*\tilde{U}
		      +\tilde{m}^2_D\tilde{D}^*\tilde{D}
		     +(h_b A_b\Phi^\dagger_1\tilde{Q}\tilde{D}
		      - h_t A_t\Phi^T_2i\tau_2\tilde{Q}\tilde{U}+{\rm H.c.})
                      \,, \\
-{\cal L}_F &=& |h_b\Phi^\dagger_1\tilde{Q}|^2
               +|h_t\Phi^T_2i\tau\tilde{Q}|^2
	       -(\mu h^*_b\tilde{Q}^\dagger\Phi_2\tilde{D}^*
	       +\mu h^*_t\tilde{Q}^\dagger i\tau_2\Phi^*_1\tilde{U}^*
	        +{\rm h.c.}) \nonumber\\
	    && -(h^*_b\tilde{D}^*\Phi^T_1 i\tau_2
	       + h^*_t\tilde{U}^*\Phi^\dagger_2)
	     (h_b i\tau_2\Phi^*_1\tilde{D}-h_t\Phi_2\tilde{U})\,,\\
-{\cal L}_D &=& \frac{g^2}{4}\left[2|\Phi^T_1 i\tau_2\tilde{Q}|^2
               +2|\Phi^\dagger_2\tilde{Q}|^2
	     -\tilde{Q}^\dagger\tilde{Q}(\Phi^\dagger_1\Phi_1
	        +\Phi^\dagger_2\Phi_2)\right]\nonumber\\
	    && +\frac{g^{\prime 2}}{4}(\Phi^\dagger_2\Phi_2
	        -\Phi^\dagger_1\Phi_1)
           \left[\frac{1}{3}
		(\tilde{Q}^\dagger\tilde{Q})
		-\frac{4}{3}(\tilde{U}^*\tilde{U})
		+\frac{2}{3}(\tilde{D}^*\tilde{D})\right]\,,\\
-{\cal L}_f &=& h_b\left[\bar{b}_R(t_L,b_L)^T\Phi^*_1+{\rm H.c.}\right]
	     +h_t\left[\bar{t}_R(t_L,b_L)^Ti\tau_2\Phi_2
	         +{\rm H.c.}\right]\,,
\end{eqnarray}
with the notations $\tilde{Q}^T=(\tilde{t}_L,\tilde{b}_L)^T$, 
$\tilde{U}^*=\tilde{t}_R$, and $\tilde{D}^*=\tilde{b}_R$ for the
scalar top and bottom quarks. 

The MSSM introduces several new parameters in the theory that are absent
from the SM and could, in principle, possess many $CP$--odd phases \cite{MS}. 
Specifically, the new $CP$ phases may come from the following parameters: 
(i) the higgsino
mass parameter $\mu$, which involves the bilinear mixing of the two Higgs
chiral superfields in the superpotential; (ii) the soft SUSY--breaking
gaugino masses $M_a$ ($a=1,2,3$), where the index $a$ stands for the
gauge groups U(1)$_Y$, SU(2)$_L$ and SU(3)$_c$, respectively;
(iii) the soft bilinear Higgs mixing masses $m^2_{12}$, which is
sometimes denoted as $B\mu$ in the literature; and (iv) the soft trilinear 
Yukawa couplings $A_f$ of the Higgs particles to scalar fermions. 
If the universality condition is imposed on all gaugino masses at the
unification scale $M_X$, the $M_a$'s have a common phase. Likewise, the 
different trilinear couplings $A_f$ are all equal at $M_X$, i.e.
$A_f=A$. 

The conformal--invariant part of the supersymmetric Lagrangian has two 
global U(1) symmetries; the U(1)$_Q$ Peccei--Quinn symmetry
and the U(1)$_R$ symmetry acting on the Grassmann--valued coordinates.
As a consequence, not all phases of the four complex parameters 
$\{\mu,m^2_{12},M_a,A\}$ turn out to be physical, i.e. two phases may be
removed by redefining the fields accordingly. Employing the two global 
symmetries, one of the Higgs doublets and the gaugino
fields can be re-phased such that $M_a$ and $m^2_{12}$ become real.
In this case, arg($\mu$) and arg($A$) are the only physical
$CP$--violating phases at low energies in the MSSM supplemented 
by universal boundary conditions.

We work in the theoretical framework provided by the MSSM, including 
explicitly the $CP$--violating phases $\Phi_\mu={\rm arg}(\mu)$ and
$\Phi_A={\rm arg}(A)$ of the universal trilinear parameter.
Moreover we take the case where the first and second sfermions are 
so heavy as to avoid the stringent EDM constraints and assume 
the universal soft--breaking mass parameters and phases for the third 
generation squarks:
\begin{eqnarray}
|A_t|=|A_b|\equiv A\,, \qquad
\Phi_{A_t}=\Phi_{A_b}=\Phi_A\,.
\end{eqnarray}
Consequently, all the physical quantities are determined by 
$\{\tan\beta,|A|,|\mu|,\Phi_A,\Phi_\mu\}$, and two sfermion masses as well
as the charged Higgs--boson mass $m_{H^\pm}$, as shown later.

%
%

\subsection{Neutral Higgs--boson mixing}

The most general $CP$--violating Higgs potential of the
MSSM can conveniently be described by the effective Lagrangian \cite{PW}:
\begin{eqnarray}
{\cal L}_V &=& \mu^2_1(\Phi^\dagger_1\Phi_1)
+\mu^2_1(\Phi^\dagger_2\Phi_2)
+m^2_{12}(\Phi^\dagger_1\Phi_2)
+m^{*2}_{12}(\Phi^\dagger_2\Phi_1)\nonumber\\
&& +\lambda_1(\Phi^\dagger_1\Phi_1)^2
+\lambda_2(\Phi^\dagger_2\Phi_2)^2
+\lambda_3(\Phi^\dagger_1\Phi_1)(\Phi^\dagger_2\Phi_2)
+\lambda_4(\Phi^\dagger_1\Phi_2)(\Phi^\dagger_2\Phi_1)
\nonumber\\
&&  +\lambda_5(\Phi^\dagger_1\Phi_2)^2
+\lambda^*_5(\Phi^\dagger_2\Phi_1)^2
+\lambda_6(\Phi^\dagger_1\Phi_1)(\Phi^\dagger_1\Phi_2)
+\lambda^*_6(\Phi^\dagger_1\Phi_1)(\Phi^\dagger_2\Phi_1)
\nonumber\\
&& +\lambda_7(\Phi^\dagger_2\Phi_2)(\Phi^\dagger_1\Phi_2)
+\lambda^*_7(\Phi^\dagger_2\Phi_2)(\Phi^\dagger_2\Phi_1)\,.
\end{eqnarray}
At the tree level, the kinematic mass parameters are given by
$\mu^2_1=-m^2_1-|\mu|^2$ and $\mu^2_2=-m^2_2-|\mu|^2$, and the 
quartic couplings determined by the SM gauge couplings as
\begin{eqnarray}
&& \lambda_1=\lambda_2=-\frac{1}{8}(g^2+g^{\prime 2})\,,\ \
   \lambda_3 = -\frac{1}{4}(g^2-g^{\prime 2})\,,\nonumber\\
&& \lambda_4 =\frac{1}{2}g^2\,,\qquad
   \lambda_5=\lambda_6=\lambda_7=0\,.
\end{eqnarray}
Here, $m^2_1$, $m^2_2$, and $m^2_{12}$ are soft--SUSY--breaking parameters
related to the Higgs sector. Beyond the Born approximation, the quartic
couplings $\lambda_5$, $\lambda_6$, $\lambda_7$ receive significant radiative
corrections from trilinear Yukawa couplings of the Higgs fields to 
scalar--top and scalar--bottom quarks. These parameters are in general
complex and the analytic expressions of the quartic couplings can be
found in the Appendix of Ref.~\cite{PW}.

In the existence of the $CP$--violating quartic couplings, the three neutral,
physical Higgs bosons may strongly mix with one another. 
To describe this $CP$--violating mixing, it is necessary to 
determine the ground state of the Higgs potential to 
determine physical Higgs states and their self--interactions. To this end
we introduce the linear decompositions of the Higgs fields
\begin{eqnarray}
\Phi_1=\left(\begin{array}{cc}
\phi^+_1 \\
\frac{1}{\sqrt{2}}(v_1+\phi_1+ia_1)
\end{array}\right)\,, \qquad
\Phi_2={\rm e}^{i\xi}\left(\begin{array}{cc}
\phi^+_2 \\
\frac{1}{\sqrt{2}}(v_2+\phi_2+ia_2)
\end{array}\right)\,,
\end{eqnarray}
with $v_1$ and $v_2$ the moduli of the vacuum expectation values (VEVs) of the
Higgs doublets and their relative phase $\xi$. These VEVs and the relative 
phase can be determined by the minimization conditions on ${\cal L}_V$.
It is always guaranteed that one combination of the $CP$--odd  Higgs fields
$a_1$ and $a_2$ ($G^0=\cos\beta a_1+\sin\beta a_2$) defines a flat direction
in the Higgs potential and it is absorbed as the longitudinal component of the
$Z$ field. As a result, there exist one charged Higgs state and 
three neutral Higgs states that are mixed in the presence of $CP$ violation in
the Higgs sector. Denoting the remaining $CP$--odd state $a=-\sin\beta a_1
+\cos\beta a_2$, the neutral Higgs--boson mass matrix describing the 
mixing between $CP$--even and $CP$--odd fields is given by
\begin{eqnarray}
{\cal M}^2_N=\left(\begin{array}{cc}
{\cal M}^2_P     &   {\cal M}^2_{PS} \\
{\cal M}^2_{SP}  &   {\cal M}^2_S 
\end{array}\right)\,,
\end{eqnarray}
where ${\cal M}^2_P$ and ${\cal M}^2_S$ describe the $CP$--preserving
transitions $a\rightarrow a$ and $(\phi_1,\phi_2)\rightarrow (\phi_1,\phi_2)$,
respectively, and ${\cal M}^2_{PS}=({\cal M}^2_{SP})^T$ contains the
$CP$--violating mixing $a\leftrightarrow (\phi_1,\phi_2)$. The analytic form 
of the sub-matrices is given by
\begin{eqnarray}
&&{\cal M}^2_P=m^2_a=\frac{1}{s_\beta c_\beta}\left\{
{\cal R}(m^2_{12}{\rm e}^{i\xi})
+v^2\bigg[2{\cal R}(\lambda_5{\rm e}^{2i\xi})s_\beta c_\beta
+\frac{1}{2}{\cal R}(\lambda_6{\rm e}^{i\xi})c^2_\beta
+\frac{1}{2}{\cal R}(\lambda_7{\rm e}^{i\xi})s^2_\beta\bigg]
\right\}\,, \nonumber\\
&&{\cal M}^2_{SP}=v^2\left(\begin{array}{c}
{\cal I}(\lambda_5{\rm e}^{2i\xi})s_\beta
+{\cal I}(\lambda_6{\rm e}^{i\xi})c_\beta\\
{\cal I}(\lambda_5{\rm e}^{2i\xi})c_\beta
+{\cal I}(\lambda_7{\rm e}^{i\xi})s_\beta
\end{array}\right)\,, \nonumber\\
&&{\cal M}^2_S=m^2_a\left(\begin{array}{cc}
s^2_\beta        & -s_\beta c_\beta\\
-s_\beta c_\beta & c^2_\beta
\end{array}\right)\nonumber\\
&& -\left(\begin{array}{cc}
2\lambda_1 c^2_\beta+2{\cal R}(\lambda_5{\rm e}^{2i\xi})s^2_\beta
+2{\cal R}(\lambda_6{\rm e}^{i\xi})s_\beta c_\beta &
\lambda_{34}s_\beta c_\beta+{\cal R}(\lambda_6{\rm e}^{i\xi})c^2_\beta
+{\cal R}(\lambda_7{\rm e}^{i\xi})s^2_\beta \\ 
\lambda_{34}s_\beta c_\beta+{\cal R}(\lambda_6{\rm e}^{i\xi})c^2_\beta
+{\cal R}(\lambda_7{\rm e}^{i\xi})s^2_\beta &
2\lambda_2 s^2_\beta+2{\cal R}(\lambda_5{\rm e}^{2i\xi})c^2_\beta
+2{\cal R}(\lambda_7{\rm e}^{i\xi})s_\beta c_\beta 
\end{array}\right)\,.
\end{eqnarray}
Correspondingly, the charged Higgs-boson mass is given by
\begin{eqnarray}
m^2_{H^\pm}=m^2_a+\frac{1}{2}\lambda_4 v^2
           -{\cal R}(\lambda_5{\rm e}^{2i\xi}) v^2\,.
\end{eqnarray}
Taking this very last relation and the universality of
the trilinear parameters into account, we can express the neutral
Higgs--boson masses as functions of $\{m_{H^\pm}, |\mu|, |A|,
\Phi_\mu, \Phi_{A}\}$, the common SUSY scale $M_{\rm SUSY}$, 
the parameter $\tan\beta$, and the physical induced phase $\xi$. 
However, with the chargino and neutralino contributions
neglected, the radiatively induced phase $\xi$ accompanies the 
the $\mu$ parameter without exception so that only the sum of the three
phases, $\Phi_{A\mu}=\Phi_A+\Phi_\mu+\xi$, enters the loop--corrected
Higgs potential.

The $CP$--violating induced relative phase $\xi$ between two Higgs 
doublets can be obtained analytically by combining the two relations 
for the minimization \cite{DEMIR}.
First of all, we make use of the fact that a U(1)$_{PQ}$ rotation allows
us to take $m^2_{12}$ to be real and for a notational convenience define 
$\tilde{\lambda}_6$, $\delta$, and $\delta'$;
\begin{eqnarray}
&& \tilde{\lambda}_6=\lambda_6\, c^2_\beta+\lambda_7\, s^2_\beta\,,\nonumber\\
&& \delta = \left(\frac{m^2_{H^\pm}}{v^2}-\frac{\lambda_4}{2}
                 +\lambda_5\right)\sin 2\beta\,, \nonumber\\
&& \delta'= \left(\frac{m^2_{H^\pm}}{v^2}-\frac{\lambda_4}{2}
                 -\lambda_5\right)\sin 2\beta\,. 
\end{eqnarray}
Then, the CP--violating induced phase $\xi$ is determined by the relations;
\begin{eqnarray}
&& \sin\xi = -\frac{1}{|\delta|^2}\left\{ 
              {\cal R}(\delta) {\cal I}(\tilde{\lambda}_6) 
             -{\cal I}(\delta)\sqrt{|\delta|^2-{\cal I}^2(\tilde{\lambda}_6)}
                                  \right\}\,, \nonumber\\
&& \cos\xi = +\frac{1}{|\delta|^2}\left\{ 
              {\cal I}(\delta) {\cal I}(\tilde{\lambda}_6) 
             +{\cal R}(\delta)\sqrt{|\delta|^2-{\cal I}^2(\tilde{\lambda}_6)}
                                  \right\}\,, 
\end{eqnarray}
and the soft--breaking positive bilinear mass squared $m^2_{12}$ is given by
\begin{eqnarray}
m^2_{12}=\frac{v^2}{2|\delta|^2}\left\{
         {\cal I}(\delta\delta'){\cal I}(\tilde{\lambda}_6)
        +{\cal R}(\delta\delta')\sqrt{|\delta|^2
	         -{\cal I}^2(\tilde{\lambda}_6)}
        -|\delta|^2{\cal R}(\tilde{\lambda}_6)\right\}
       \,.
\end{eqnarray}
The induced phase $\xi$ involves only ${\rm arg}(A\mu)$, i.e. 
$\Phi=\Phi_A+\Phi_\mu$ and so vanishes when $\Phi$ vanishes, even if
each of the CP--violating phases might not have to vanish. 
On the other hand, the size of $\delta$ or $\delta'$ is proportional to  
the pseudoscalar mass $m_a$ to a very good approximation so that if 
large, i.e, decoupled, the induced phase $\xi$ is diminished. Since the 
size of the induced phase is also inversely proportional to $\sin 2\beta$, 
the phase grows with increasing $\tan\beta$. 

Clearly, the $CP$--even and $CP$--odd states mix unless all the imaginary
parts of the parameters $\lambda_5,\lambda_6,\lambda_7$ are vanishing. 
Since the Higgs---boson mass matrix ${\cal M}^2_N$ describing the mixing is 
symmetric, we can diagonalize it by means of an orthogonal rotation $O$;
\begin{eqnarray}
O^T{\cal M}^2_N O={\rm diag}(m^2_{H_3},m^2_{H_2},m^2_{H_1})\,,
\end{eqnarray}
with the ordering $m_{H_1}\leq m_{H_2}\leq m_{H_3}$.
The size of the resulting
$CP$--violating neutral Higgs--boson mixing  is characterized 
by the factor $(1/32\pi^2)(|\mu||A_f|Y_f^4/M^2_{\rm SUSY})\sin\Phi_{A_f\mu}$, 
where $Y_f$ is the Yukawa coupling of the fermion $f$, $\Phi_{A_f\mu}={\rm
Arg}(A_f\mu)+\xi$ for $f=t,b$, and $M_{\rm SUSY}$ is a typical
SUSY--breaking scale, which might be taken to be the average of the
two sfermion ($\tilde{f}_1, \tilde{f}_2$) masses squared.

\subsection{Sfermion mixing}

The sfermion mass matrix squared of the left-/right-handed
sfermions is given by
\begin{eqnarray}
{\cal M}^2_{\tilde{f}}=\left(\begin{array}{cc}
          A_f                &  B_f {\rm e}^{-i\phi_f}   \\
      B_f {\rm e}^{i\phi_f}  &        C_f 
                           \end{array}\right)\,,
\end{eqnarray}
where, for the case of scalar top quarks,  
\begin{eqnarray}
A_t &=& m^2_{\tilde{t}_L} +m^2_t
       +\frac{1}{6}(4m^2_W-m^2_Z)\cos 2\beta\,,\nonumber\\
B_t &=& m_t|A^*_t-\mu {\rm e}^{i\xi}\cot\beta|\,,\nonumber\\
C_t &=& m^2_{\tilde{t}_R}+m^2_t+\frac{2}{3}m^2_Zs^2_W\cos 2\beta\,, \nonumber\\
\phi_t &=& {\rm arg}(A_t-\mu^* {\rm e}^{-i\xi}\cot\beta)\,,
\end{eqnarray}
and, for the case of scalar bottom quarks,
\begin{eqnarray}
A_b&=&m^2_{\tilde{b}_L} +m^2_b
       -\frac{1}{6}(2m^2_W+m^2_Z)\cos 2\beta\,,\nonumber\\
B_b&=&m_b|A^*_b-\mu {\rm e}^{i\xi}\tan\beta|\,,\nonumber\\
C_b&=& m^2_{\tilde{b}_R} +m^2_b-\frac{1}{3}m^2_Zs^2_W\cos 2\beta\,, \nonumber\\
\phi_b&=& {\rm arg}(A_b-\mu^* {\rm e}^{-i\xi}\tan\beta)\,,
\end{eqnarray}
where $m^2_{\tilde{t}_{L,R}}$ and $m^2_{\tilde{b}_{L,R}}$ are the
left/right--handed soft--SUSY--breaking stop/sbottom quark masses
squared, respectively.
The Hermitian mass matrix squared ${\cal M}^2_{\tilde{f}}$ can be 
diagonalized by an unitary transformation
\begin{eqnarray}
U^\dagger_f\,{\cal M}^2_{\tilde f} \, U_f
  ={\rm diag}(m^2_{\tilde{f}_1}, m^2_{\tilde{f}_2})
  \qquad [\,m_{\tilde{f}_1}\leq m_{\tilde{f}_2}\,]\,,
\end{eqnarray}
with the parameterization for the unitary matrix 
\begin{eqnarray}
U_f =\left(\begin{array}{cc}
     \cos\theta_f               & -\sin\theta_f\, {\rm e}^{-i\phi_f} \\
 \sin\theta_f\, {\rm e}^{i\phi_f} &  \cos\theta_f
           \end{array}\right)\,,
\end{eqnarray}
satisfying $-\pi/2\leq \theta_f \leq 0$, and the sfermion mass 
eigenvalues and mixing angles are given by
\begin{eqnarray}
&& m^2_{\tilde{f}_{1,2}}=M^2_{\rm SUSY}\mp\frac{1}{2}\Delta_f\,,\nonumber\\
&& \sin2\theta_f =-\frac{2B_f}{\sqrt{(A_f-C_f)^2+4B^2_f}}\,,\nonumber\\
&& \cos2\theta_f =-\frac{A_f-C_f}{\sqrt{(A_f-C_f)^2+4B^2_f}}\,,
\end{eqnarray}
with the phenomenological parameters $M^2_{\rm SUSY}$
and $\Delta_f$ defined as
\begin{eqnarray}
&& M^2_{\rm SUSY}=\frac{m^2_{\tilde{f}_2}+m^2_{\tilde{f}_1}}{2}=
                  \frac{A_f+C_f}{2}\,, 
   \nonumber\\
&& \Delta_f=m^2_{\tilde{f}_2}-m^2_{\tilde{f}_1}=\sqrt{(A_f-C_f)^2+4B^2_f}\,.
\end{eqnarray}
We note that there exists a freedom of redefining the sfermion fields
by phase rotations. So, in constructing the Feynman rules for 
the Higgs--sfermion--sfermion vertices, we make the transformation 
of the sfermion basis $(\tilde{f}_L, \tilde{f}_R)$ into the mass basis 
$(\tilde{f}_1,\tilde{f}_2)$ by use of the unitary matrix $U_f$
and a $2\times 2$ phase rotation matrix $P_f$:
\begin{eqnarray}
\left(\begin{array}{c}
        \tilde{f}_L \\
	\tilde{f}_R
      \end{array}\right)
  =
P_f\, U_f \left(\begin{array}{c}
            \tilde{f}_1 \\
	    \tilde{f}_2 
	  \end{array}\right)\,,
\end{eqnarray}
where the $2\times 2$ matrix $P_f$ is defined in terms of the
phase of the trilinear term $A_f$ as 
\begin{eqnarray}
P_f = {\rm diag}\left(\,{\rm e}^{-i\Phi_{A_f}/2},\ \ 
                      {\rm e}^{i\Phi_{A_f}/2}\,\right)\,.
\end{eqnarray}
The additional phase rotation renders the Higgs--sfermion--sfermion 
couplings dependent only on the sum of the phases, $\Phi_{A_f\mu}=\Phi_{A_f}
+\Phi_\mu +\xi$. Furthermore, the orthogonal matrix $O$ transforming 
the Higgs boson weak basis into the Higgs boson mass basis involves only 
the sum of the phases as noted before. As a result, with the universal 
trilinear term assumed,  only one combination 
$\Phi_{A\mu}=\Phi_A+\Phi_\mu+\xi$ of the $CP$--violating 
phases dictates all the physical quantities related to the sfermion sectors 
and the $CP$--violating induced neutral Higgs--boson mixing.

Not only the phase $\Phi_{A\mu}$ but also the magnitudes of $|A|$ and $|\mu|$ 
affect the values of the sfermion masses by modifying the mass
splitting. Since we are interested mainly in
the effects of the $CP$--violating phases, we take the real parameters 
to be fixed. In this case, the allowed range for the 
$CP$--phase $\Phi_{A\mu}$ is determined by the values of the real parameters 
through the constraint:
\begin{eqnarray}
\Delta \leq M^2_{\rm SUSY}\,. 
\label{eq:tighter}
\end{eqnarray}
In principle, $\Delta$ can be as large as $2M^2_{\rm SUSY}$, but
the expressions of loop corrections to the Higgs potential \cite{PW} 
is valid when the tighter constraint (\ref{eq:tighter}) is satisfied. 
With a simple assumption that two diagonal elements of the stop mass matrix,
$A_t$ and $C_t$, are equal, the constraint (\ref{eq:tighter}) leads to 
the inequality for the phase $\Phi_{A\mu}$:
\begin{eqnarray}
\cos\Phi_{A\mu}\geq\frac{|A|^2+|\mu|^2/\tan^2\beta
                          -\frac{M^4_{\rm SUSY}}{4m^2_t}}{
                     2|A||\mu|/\tan\beta}\,,
\end{eqnarray}
showing that the $CP$--odd phase $\Phi_{A\mu}$ could be constrained by 
the measurements
of the scalar top--quark masses, depending on the values of real parameters
$|A|$ and $|\mu|$.

Fig.~2 exhibits the allowed maximal value of the $CP$--violating phase 
$\Phi_{A\mu}$ on the plane of $\{|A|,|\mu|/\tan\beta\}$ for the scalar 
top--quark
sector for $M_{\rm SUSY}=500$ GeV and $m_t=175$ GeV. A few
comments are in order: (i)  the full range of the phase is not allowed unless 
$|A|$ or $|\mu|/\tan\beta$ is less than $M^2_{\rm SUSY}/2m_t\approx 710$ GeV;
(ii) the upper-left and lower-right regions are not 
allowed due to the self--evident constraint $|\cos\Phi_{A\mu}|\leq 1$; 
(iii) the lower-left hatched triangular area allows the full range for 
the phase $\Phi_{A\mu}$; (iv) if $A_t$ is not equal to $C_t$, the phase is
more constrained. In addition, the fact that every physical quantity
is determined by only the combined phase $\Phi_{A\mu}=\Phi_A+\Phi_\mu+\xi$
implies that large $CP$--violating effects are caused when 
$|A|$ and $|\mu|/\tan\beta$ are similar as well as large in size. 
One natural consequence is then that for a large $\tan\beta$, which is
disfavored by the two--loop EDM limits \cite{CKP}, the effects of the
$CP$ phases are reduced. Based on these observations, we consider 
two sets of $\{|A|, |\mu|\}$ for 
a fixed value of $\tan\beta=3$ and $M_{\rm SUSY}=500$ GeV in the following 
numerical analysis
\footnote{
For large values of $A_t-\mu/\tan\beta$ and moderate $m_{{\tilde t}_1}$,
$\Gamma(gg\rightarrow H)$ can be siginificantly reduced by even an order of
magnitude due to the destructive stop contributions \cite{KKD}. 
This destructive interference effect is not so large in our choice
of SUSY parameters.
} :
\begin{eqnarray}
RR1&:& \{|\mu|=0.5\,{\rm TeV}\,, |A|= 0.5\,{\rm TeV}\}\,,\nonumber\\ 
RR2&:& \{|\mu|=1.2\,{\rm TeV}\,, |A|= 0.4\,{\rm TeV}\}\,.
\end{eqnarray}

We note that the full range for the phase $\Phi_{A\mu}$ is allowed in
the $RR1$, but the phase is restricted to the intervals $[0^0,140^0]$ 
and $[220^0,360^0]$ in the $RR2$. The Higgs--boson production rates,
which we are interested in, are $CP$--even observables and so they
are dependent on only the cosine of the phase $\Phi_{A\mu}$. 
Keeping in mind this point we have considered only the range $[0^0,180^0]$ 
for the phase $\Phi_{A\mu}$ in Fig.~2.

Experimentally, the tightest limits on the squark masses come 
from direct searches at the Tevatron. Concerning the
$\tilde{t}_1$ mass, the limit on $m_{\tilde{t}_1}$ can be drawn at 
around 120 GeV or so for a large value of $\tan\beta$
assuming the mass of the lightest neutralino is smaller than 50 GeV
\cite{Stuart}.
A less model--dependent mass bound of 90 GeV is obtained at LEP2
\cite{OPAL-SQ}.
As for the lightest sbottom mass, $m_{\tilde{b}_1}$,
this is excluded for somewhat lower values \cite{D0}. 
Since both scenarios
give the lightest stop mass larger than 200 GeV, 
the present experimental bounds on the sfermion masses can be safely
neglected in our analysis.

\section{Higgs--Boson Production at Hadron Colliders}
\label{sec:production}

\subsection{Parton--level production cross section}

The basic production processes of supersymmetric Higgs particles at 
hadron colliders are essentially the same as in the Standard Model (SM).
Important differences are nevertheless generated by the modified couplings,
the extended particle spectrum, and the existence of Higgs bosons of mixed
parity. There are three crucial contributions to the Higgs--boson
production in gluon--gluon collisions; (i) heavy fermion contributions,
(ii) heavy sfermion contributions affected by sfermion mixing due to the 
large Yukawa couplings, and (iii) large Higgs--boson mixing due to the same
large Yukawa couplings. The contributions (i) and (ii) are at the one--loop
order, while the last contribution (iii) is at the two--loop order in
character. However, the one--loop Higgs--boson mixing cannot be neglected
as a higher--loop effect, because even a small loop contribution
to the scalar--pseudoscalar mixing term can cause a maximal mixing 
between the heavier $CP$--even and $CP$--odd neutral Higgs fields due to their
intrinsic close degeneracy in the MSSM.

The leading--order (LO) partonic cross section 
$\sigma_{LO}(gg\rightarrow H_i)$ $[i=1,2,3]$ 
for the gluon fusion of Higgs particles in the MSSM with explicit
$CP$ violation can be expressed in the narrow--width approximation
by their scalar and pseudoscalar couplings,
$g^i_{sff}$ and $g^i_{pff}$ to each fermion $f$, and their couplings
$g^i_{\tilde{f}\tilde{f}}$, to each sfermion $\tilde{f}$:
\begin{eqnarray}
&& \sigma_{LO}(gg\rightarrow H_i)=\hat{\sigma}_{LO}
      \,\delta\left(1-\frac{m^2_{H_i}}{\hat{s}}\right)\,,\\ 
&& \hat{\sigma}_{LO}=\frac{\alpha^2_s(Q)}{256\pi}
    \left\{\bigg|\sum_f\frac{g^i_{sff}}{m_f} A_{sf}(\tau_{if})
     +\frac{1}{4}\sum_{\tilde{f}}
     \frac{g^i_{\tilde{f}\tilde{f}}}{m^2_{\tilde{f}}}
      A_{\tilde{f}}(\tau_{i\tilde{f}})\bigg|^2
     +\bigg|\sum_f\frac{g^i_{pff}}{m_f} A_{pf}(\tau_{if})\bigg|^2
     \right\} \,,
\end{eqnarray}
where the explicit form of the couplings $\{g^i_{sff},
g^i_{pff},g^i_{\tilde{f}\tilde{f}}\}$ is given in the Appendix and
the form factors $A_{sf}$, $A_{pf}$ and $A_{\tilde{f}}$ can be
expressed in terms of the scaling function $f(\tau_{ix}=m^2_{H_i}/4m^2_x)$:
\begin{eqnarray}
A_{sf}(\tau)&=&\tau^{-1}\,[1+(1-\tau^{-1}) f(\tau)]\,,\nonumber\\
A_{pf}(\tau)&=&\tau^{-1}\,f(\tau)\,,\nonumber\\
A_{\tilde{f}}(\tau)&=&\tau^{-1}\,[-1+\tau^{-1}f(\tau)]\,,
\end{eqnarray}
where the scaling function $f(\tau)$ stands for the integrated function
\begin{eqnarray}
f(\tau)=-\frac{1}{2}\int_0^1\frac{{\rm d}y}{y}
\ln\left[1-4\tau y(1-y)\right]
=\left\{\begin{array}{cl}
{\rm arcsin}^2(\sqrt{\tau}) \,,   &\qquad \tau\leq 1\,, \\
-\frac{1}{4}\left[\ln \left(\frac{1+\sqrt{1-\tau^{-1}}}{
1-\sqrt{1-\tau^{-1}}}\right)
-i\pi\right]^2\,, & \qquad \tau\geq 1\,.
\end{array}\right.
\end{eqnarray}
For small $\tan\beta$ the contribution of the top loop is 
dominant, while for large $\tan\beta$ the bottom loop is
strongly enhanced.  The contributions of the squark loops can be significant 
for the Higgs-boson masses above the squark--pair thresholds \cite{DDS}.

The limits of both large and small loop masses are interesting for 
SUSY Higgs particles. The contribution of the top loop to the
$H_1gg$ coupling can be calculated approximately in the limit of large
loop masses, while the bottom contributions to the $H_i gg$ couplings
can be calculated in the approximation of small $b$ masses.
The limits of large loop masses are approximated with the following
limiting values of the form factors,
\begin{eqnarray}
A_{sf} \rightarrow 2/3 \,,\qquad 
A_{pf} \rightarrow 1  \,, \qquad
A_{\tilde{f}} \rightarrow 1/3\,,
\end{eqnarray}
and the $A_{pf}$ is not altered by QCD radiative corrections due to the
non--renormalization of the axial anomaly. Incidentally, $g^i_{sff}$ and 
$g^i_{pff}$ are always proportional to the Yukawa coupling $Y_f$, i.e.
the fermion mass $m_f$ so that the fermionic contributions remain finite
even for large fermion masses. However, the sfermion--loop contributions
vanish because of the suppression by the factor $1/m^2_{\tilde{f}}$.
On the other hand, in the opposite limit where the quark--loop
mass is much smaller than the Higgs masses, the amplitudes are the same
for the scalar and pseudoscalar form factors:
\begin{eqnarray}
A_{sf}= A_{pf} \rightarrow -\frac{1}{4\tau}
       \bigg[\ln(4\tau)+i\pi\bigg]^2\,.
\end{eqnarray}
This result follows from the restoration of chiral symmetry in the limit
of vanishing quark masses.

If the mass of the produced Higgs boson is smaller than the $t\bar{t}$ and
$\tilde{t}_1\tilde{t}_1^*$ pair thresholds, the form factors are real;
above the thresholds they are complex. Fig.~3 shows the dependence of the
real (solid) and imaginary (dashed) parts of the form factors, 
$A_{sf}$ (left), $A_{pf}$ (middle), and $A_{\tilde{f}}$ (right) on 
the Drell--Yan variable $\tau$. All the real parts are peaked at 
$\tau=1$ above which all the imaginary parts become finite.
Since the fermionic form factors are determined by the top quark mass
$m_t=175$ GeV, the production cross sections are expected to be
enhanced near $m_H=2m_t$ with unsuppressed couplings of the Higgs boson 
to top quarks. Similarly, the sfermionic form factor 
will be enhanced for $m_H=2 m_{\tilde{t}}$ with unsuppressed couplings
of the Higgs boson to scalar top quarks.

The Higgs bosons and sparticles of the MSSM that enter the process
$gg\rightarrow H_i$ can be also produced 
via other channels, both as real and virtual objects. 
{}From their search at past and present colliders, several limits on their
masses and couplings have been drawn. As for the neutral Higgs bosons
of the MSSM, the most stringent bounds come from LEP in the $CP$--invariant
theories. 
For both $m_{h^0}$ and $m_{A^0}$ these are set at around 80
GeV by all the LEP Collaborations \cite{ADLO} for $\tan\beta >1$.
There is also a more recent and somewhat higher limit on $m_{h^0}$ 
from ALEPH \cite{ALEPH} 
of about $85$ GeV 
\footnote{
The most recent limit obtained by the LEP collaborations at the end of
the '99 run which is still preliminary and not combined is
90 GeV \cite{PRE-MH}.  } 
for $\tan\beta\geq 1$ at 95\% confidence
level, using data collected at the c.m. energies in the range between 
192 GeV and 196 GeV and a total luminosity of about 100 pb$^{-1}$.
However, the existence of $CP$--violating phases could weaken the present 
experimental bounds on the lightest Higgs mass up to about 60 GeV \cite{PW}.

\subsection{Hadronic level production cross section}

In the $CP$--preserving limit, the cross section  
near the top--quark--pair threshold are not reliable taking
into account the perturbative two--loop QCD corrections which are infinite at
the threshold \cite{DDS}.

It is well known \cite{DDS,Dawson} that next--to--leading order 
(NLO) corrections to the processes $gg\rightarrow H_i$ from ordinary QCD 
are very large. However, it has been shown that they affect the
quark and squark contributions very similarly \cite{DDS}, and
can be approximated by the so--called 
$K$ factors, defined by the ratios of the higher--order cross sections 
to the leading order cross sections. They vary little with the masses
of the neutral Higgs bosons in general, yet they depend strongly 
on $\tan\beta$. For small $\tan\beta$, their size is between 1.5 and 1.7; 
for large $\tan\beta$ they are in general close to unity, except when
the lightest Higgs boson approaches the SM domain. Therefore,  
we may take $K$ to be a constant for a given $\tan\beta$ to a good 
approximation
\footnote{
Since the QCD correction to the $CP$-odd part of the production amplitude
is singular at the top--quark--pair threshold, the cross sections are not
realiable near the threshold \cite{MSY}.
}.

In order to find the physical cross section it is necessary to 
integrate the parton--level production cross section with the 
distribution of gluons in a proton,
\begin{eqnarray}
\sigma(pp\rightarrow H_i) &=& K \int {\rm d}x_1{\rm d}x_2\,
    g(x_1,Q) g(x_2,Q)
    \, \sigma_{LO}(gg\rightarrow H_i)\nonumber\\
    &=& K\, \hat{\sigma}_{LO}(gg\rightarrow H_i)\,\, \tau 
     \frac{{\rm d}{\cal L}^{gg}_{LO}}{{\rm d}\tau} \,,
\end{eqnarray}
where $g(x,Q)$ is the distribution of gluons in the proton and the
the gluon--gluon luminosity is defined as
\begin{eqnarray}
\tau \frac{{\rm d}{\cal L}^{gg}}{{\rm d}\tau}=\int^1_\tau {\rm d}x 
     \,\frac{\tau}{x}\, g(x,Q)\, g(\tau/x,Q)\,,
\end{eqnarray}
and the Drell--Yan variable $\tau =m^2_{H_i}/s$
with $s$ being the invariant hadron collider energy squared.  
We note that the cross section decreases with increasing Higgs mass.
This is, to a large extent, a consequence of the sharply--falling
$gg$ luminosity for large Higgs--boson masses. 
Therefore, the significant modifications of the lightest Higgs mass by
the radiative corrections cannot be neglected.

The dependence of the cross sections on the Higgs--boson masses
is encoded in the gluon--gluon luminosity function and the parton--level 
cross sections $\hat{\sigma}(gg\rightarrow H_i)$, the latter of which 
is expected to have a weaker dependence.  On the other hand, 
the lightest Higgs--boson mass is less than 130 GeV irrespective of the
CP--violating phases and its lower limit can be set by its present 
collider searches. This limit is of course dependent on the CP--violating
phases. Nevertheless, in our analysis we simply take 70 GeV for the
lower limit of the lightest Higgs-boson mass. Fig.~4 shows the gluon-gluon 
luminosity function $\tau ({\rm d}{\cal L}^{gg}/{\rm d}\tau)$ with respect 
to the mass of the produced Higgs mass at the Tevatron with$\sqrt{s}=2$ 
TeV (solid line) and at the LHC  with $\sqrt{s}=14$ TeV (dashed line),
taking CTEQ4m \cite{CTEQ4m} as the parton distribution routine. Note that
the luminosity function is larger at the LHC at least by two orders 
of magnitude than at the Tevatron, and  the difference is enlarged 
for large Higgs masses. So, the Higgs--boson production rates are 
much larger at the LHC and remain considerable for the Higgs--boson masses 
up to several hundred GeV, while the production rates are strongly 
suppressed at the Tevatron for large Higgs masses.

\section{Numerical Results}
\label{sec:result}

In our numerical analysis, we include the contributions from only the
loops of the $t$, $b$, $\tilde{t}_{1,2}$, and $\tilde{b}_{1,2}$,
that are indeed the dominant terms because of their large Yukawa couplings. 
We take $\tan\beta$ to be 3 and assume for the scalar top/bottom
quarks masses 
$M_{\rm SUSY}=500$ GeV and $\Delta= (500\,{\rm GeV})^2$ for which
the masses of the lighter and heavier scalar top quarks are approximately
given by
\begin{eqnarray}
m_{\tilde{t}_1}=350\,{\rm GeV}\,,\qquad
m_{\tilde{t}_2}=610\,{\rm GeV}\,,
\end{eqnarray}
respectively. Therefore, the Higgs--boson production is dominated by the 
contributions from the top quark and the lighter scalar top--quark  
loops if the Higgs--boson masses are less than 1 TeV. 

As shown in Sec.~II, the $CP$--violating phases in the sfermion sectors affect the
production rates of MSSM Higgs--bosons in two ways; through the direct
loop contributions and through the neutral Higgs-boson mixing.
In order to estimate the relative size of two contributions, firstly
we consider the ratios of the production cross sections with the
top--quark and sfermion loop corrections to those with only the top--quark
loop contributions without including the neutral Higgs--boson mixing
$O$:
\begin{eqnarray}
\frac{\hat{\sigma}^0_{LO}[f+\tilde{f}]}{\hat{\sigma}^0_{LO}[f]}\,,
\label{eq:oo}
\end{eqnarray}
where the superscript $0$ indicates that the neutral Higgs--boson
mixing is not included.
In Fig.~5 we present the ratios with respect to the neutral Higgs--boson
masses. The upper three frames are in the 
scenario $RR1$, while the lower three frames are in the scenario $RR2$. 
The phase $\Phi=\Phi_A+\Phi_\mu$ is taken to be $0^0$ (solid),
$60^0$ (dashed), $120^0$ (dotted), and $180^0$ (dot--dashed).
We note in passing that in $RR2$ the allowed maximal angle of
the phase is about $140^0$. Concerning the ratios (\ref{eq:oo})
we make a few comments:
\begin{itemize}

\item[{(a)}] The stop--loop diagrams contribute little
 to the production rates unless the Higgs--boson masses are similar to
 or larger than the stop--pair threshold.  

\item[{(b)}] In the $CP$--invariant theories, $H_2$ is the $CP$--odd 
 pseudoscalar, which does not couple to the diagonal pairs of sfermions.
 So, the ratio is a unity for $\Phi=0$ irrespective of the Higgs--boson
 mass $m_{H_2}$ as shown with the solid lines in the middle frames.

\item[{(c)}] For $H_2$, the production rates are enhanced 
 due to the appearance of the $H_2\tilde{t}_1\tilde{t}^*_1$ coupling
 for the non--trivial $CP$--violating phase $\Phi$ and the size of 
 the enhancement
 is the same for $\Phi=60^o$ and $120^o$, reflecting the fact that
 the diagonal $H_2\tilde{t}_1\tilde{t}^*_1$ coupling is proportional
 to $\sin 2\Phi$.

\item[{(d)}] The $CP$--violating phase $\Phi$ changes the ratio for $H_3$
 more significantly than that of $H_2$ above the stop--pair threshold. 
 This is because the new stop--loop diagram interferes with the top--loop
 diagram at the amplitude level only for the $CP$-even $H_3$, but
 not for the $CP$--odd $H_2$.

\item[{(e)}] A large value of $|\mu|$ enhances the sensitivity of the
 production rates of the heavy Higgs--bosons to the phase $\Phi$ 
 for $m_{H_{2,3}}\geq 2 m_{\tilde{t}_1}$, in particular, for the
 heaviest Higgs boson.

\end{itemize}
Consequently, the stop--loop contributions can manifest themselves
in the Higgs--boson production in gluon--gluon fusion only if
the Higgs--boson mass is similar to or larger than the stop--pair threshold.

Secondly, the neutral Higgs--boson mixing significantly modifies 
the couplings of the Higgs bosons to fermions as well as gauge bosons in 
the presence of the $CP$--violating phases \cite{PW}. 
Therefore, it is interesting to consider
the ratios of the Higgs--boson production cross sections with the neutral 
Higgs--boson mixing to those without the mixing, while the top loop 
contributions but not the sfermion loop corrections are included:
\begin{eqnarray}
\frac{\hat{\sigma}_{LO}[f]}{\hat{\sigma}^0_{LO}[f]}\,.
\label{eq:of}
\end{eqnarray}
Fig.~6 shows the ratios (\ref{eq:of}) with respect to the produced 
Higgs--boson masses.
The upper three frames are in the scenario $RR1$, while the lower three 
frames are in the scenario $RR2$. The phase $\Phi=\Phi_A+\Phi_\mu$ 
is taken to be $0^0$ (solid), $60^0$ (dashed), $120^0$ (dotted), 
and $180^0$ (dot--dashed), and  in $RR2$ the phase angle is allowed up to 
about $140^0$. 

The ratios (\ref{eq:of}) in Fig.~6 present several interesting aspects:
\begin{itemize}
\item[{(a)}] In both $RR1$ and $RR2$ the production rates for the lightest
 Higgs boson changes significantly with the phase $\Phi$, reflecting 
 the fact that
 the couplings of the Higgs boson to the top--quark pair is
 modified significantly through the neutral Higgs---boson mixing.
\item[{(b)}] On the contrary, the production rates for the heavy Higgs
 bosons do not change so much except for the Higgs--boson masses around the
 top--quark pair threshold. It is also worthwhile to note that
 if the ratio for $H_2$ is enhanced, the ratio for $H_2$ is suppressed, and
 vice versa. This is a reflection of the fact that for a large 
 Higgs--boson mass, 
 the heavy Higgs bosons shows a two--state mixing.
\end{itemize}
Consequently, if only the top-quark loops are considered, the neutral 
Higgs--boson mixing affects the production of
the lightest Higgs boson significantly, but it does not affect 
the production of the heavy Higgs boson significantly
unless the masses are around the top--quark pair
threshold.

Thirdly, in order to estimate the effects of the neutral Higgs--boson 
mixing to the
Higgs--boson production with both the fermion and sfermion loops,
let us consider the ratios of the full production cross sections
with the neutral Higgs--boson mixing to those without 
the neutral Higgs--boson mixing, which have been considered 
by Dedes and Moretti \cite{DeMo}:
\begin{eqnarray}
\frac{\hat{\sigma}_{LO}[f+\tilde{f}]}{\hat{\sigma}^0_{LO}[f+\tilde{f}]}\,,
\label{eq:fft}
\end{eqnarray}
In Fig.~7 we present the ratios (\ref{eq:fft}) of the Higgs--boson production 
cross sections with the neutral 
Higgs--boson mixing to those without the mixing with respect to the 
mass of the produced Higgs boson. The upper three frames are in the 
scenario $RR1$, while the lower three frames are in the scenario $RR2$. 
The phase $\Phi=\Phi_A+\Phi_\mu$ is taken to be $0^0$ (solid),
$60^0$ (dashed), $120^0$ (dotted), and $180^0$ (dot--dashed).
We note that in $RR2$ the phase angle is allowed only up to
$140^0$.

Comparing Fig.~7 with Figs.~5 and 6 we observe several interesting 
aspects:
\begin{itemize}

\item[{(a)}] On the whole, the ratios are very sensitive to the 
 $CP$--violating phase $\Phi$ for large $|A|$ and $|\mu|$, 
 in particular, for a large value of $|\mu|$.

\item[{(b)}] The production rate of the lightest Higgs boson is modified
 mainly by the neutral Higgs--boson mixing as clearly seen by comparing
 Fig.~6 and Fig.~7.

\item[{(c)}] In $RR1$, the production rates for the heavier Higgs 
 bosons are not affected by the neutral Higgs-boson mixing.  
 However, they change a lot in the scenario $RR2$ with a large value of
 $|\mu|$. The production rates are most sensitive to the $CP$--violating
 phase $\Phi$ if $M_{H_{2,3}}\geq 2 m_{\tilde{t}_1}\approx 710$ GeV, i.e. 
 the threshold for a light stop--pair. 

\item[{(d)}] Comparing Figs.~5 and 6 with Fig.~7, one can easily find
 that the couplings of the heavy Higgs bosons to sfermions also 
 are significantly modified by the neutral Higgs--boson mixing, 
 resulting in a large enhancement of $\hat{\sigma}_{LO}(gg\rightarrow H_2)$
 and a simultaneous large suppression of $\hat{\sigma}_{LO}(gg\rightarrow
 H_3)$ for non--trivial values of the phase $\Phi$.

\end{itemize}
To summarize, although the neutral Higgs--boson mixing affects 
the Higgs--boson production by gluon-gluon 
fusion at the two--loop level, their contributions to the production rates can be very significant for some
non--trivial values of the phase $\Phi$, when the higgsino mass parameter 
$|\mu|$ and the trilinear term $|A|$ are sizable. In addition, we emphasize
that the neutral Higgs--boson mixing modifies the mass spectrum of neutral
Higgs bosons as well as the couplings of the Higgs bosons to 
fermions, gauge bosons, and scalar fermions significantly. 

Finally, we consider the parton--level Higgs--boson production 
cross sections $\hat{\sigma}_{LO}(gg\rightarrow H_i)$ ($i=1,2,3$)
including the effects of the $CP$--violating phases in both the 
sfermion loops and the neutral Higgs--boson mixing.
In Fig.~8 we present the cross sections with respect to the Higgs--boson 
masses, $m_{H_1}$ (left), $m_{H_2}$ (middle), and $m_{H_3}$ (right) in 
$RR1$ (upper) and $RR2$ (lower) for four values of
$\Phi$; $0^0$ (solid), $60^0$ (dashed), $120^0$ (dotted), 
and $180^0$ (dot--dashed).

Concerning the dependence of $\hat{\sigma}_{LO}(gg\rightarrow H_i)$
on the mass $m_{H_i}$ and the phase $\Phi$, we find several interesting
aspects:
\begin{itemize}

\item[{(a)}] $\hat{\sigma}_{LO}(gg\rightarrow H_1)$
 increases sharply with increasing $m_{H_1}$ because of the sharp increase
 of $A_{sf}$ with $m_{H_1}$ for $m_{H_1}\leq 2m_t\approx 350$ GeV. 
 In $RR2$, the $CP$--violating phase could suppress the cross section 
 by a factor of five around $m_{H_1}=90$ GeV in spite of the enhancement by
 the neutral Higgs--boson mixing so that the possibility of
 finding the lightest Higgs--boson at the LHC might be seriously reduced.

\item[{(b)}] $\hat{\sigma}_{LO}(gg\rightarrow H_2)$, 
 where $H_2$ is a $CP$--odd pseudoscalar in the 
 $CP$--invariant limit, is maximal at $m_{H_2}=2m_t$, because 
 $A_{pf}$ has a sharp peak at $\tau=1$, i.e. the $t\bar{t}$ threshold.

\item[{(c)}] $\hat{\sigma}_{LO}(gg\rightarrow H_2)$
 is strongly modified by the $CP$--violating phase $\Phi$ near the 
 $\tilde{t}_1\tilde{t}^*_1$ threshold. This is because a non--trivial 
 coupling of the $CP$--odd content of $H_2$ to diagonal stop pairs
 is developed and  $H_2$ is no longer a pure $CP$ eigenstate due to 
 the neutral Higgs--boson mixing in the existence of the phase $\Phi$. 
 There is also  a little suppression in the production rates at the 
 top--pair threshold due to the suppression of the coupling of $H_2$ to 
 top quarks.

\item[{(d)}] $\hat{\sigma}_{LO}(gg\rightarrow H_3)$ has a relatively 
 dull peak at the $t\bar{t}$ threshold, as expected from the dependence 
 of $A_{sf}$ on the Drell--Yan variable $\tau$ in Fig.~4.

\item[{(e)}] The $CP$--violating phase $\Phi$ changes 
 $\hat{\sigma}_{LO}(gg\rightarrow H_3)$ significantly near the 
 $\tilde{t}_1\tilde{t}^*_1$ threshold; the enhancement is due to the
 stop--loop contributions, even though the effects are reduced by
 the neutral Higgs--boson mixing.

\end{itemize}

To recapitulate, the $CP$--violating phase $\Phi$ changes 
$\hat{\sigma}_{LO}(gg\rightarrow H_1)$ significantly  over 
the whole range of the mass $m_{H_1}$ between 70 GeV and 110 GeV.
$\hat{\sigma}_{LO}(gg\rightarrow H_2)$ is maximal at the $t\bar{t}$ 
threshold, and its size is enhanced significantly near the 
threshold of the light stop--pair by the phase $\Phi$. And, 
$\hat{\sigma}_{LO}(gg\rightarrow H_3)$ is modified significantly
at both the $t\bar{t}$ and $\tilde{t}_1\tilde{t}^*_1$ thresholds.

\section*{Conclusions}

We have investigated the effects of the $CP$--violating phases 
on the gluon--fusion process for the production of three neutral Higgs 
bosons including both the sfermion--loop (as well
as fermion--loop) contributions and the one--loop induced $CP$--violating 
scalar--pseudoscalar mixing 
in the minimal supersymmetric standard model with explicit $CP$ violation. 

We have observed two generic aspects for $CP$ violation from the stop/sbottom 
sectors:
\begin{itemize}
\item[{(a)}] If a universal trilinear parameter $|A|$ is assumed, 
  the $CP$--violating phases $\Phi_A$ and $\Phi_\mu$ enter every physical 
  quantity with only the sum $\Phi=\Phi_A+\Phi_\mu$.
\item[{(b)}] The (measured) stop masses could constrain the $CP$--violating
  phase, depending on the values of $|A|$ and $|\mu|/\tan\beta$.
\end{itemize}
{}From a detailed numerical analysis, we have found that 
the $CP$--violating phase $\Phi$ could reduce the production rate 
of $H_1$ significantly, (which is dominated by the top--quark loop) and
the phase enhances the production rate of $H_2$ and $H_3$ significantly
near and above the stop--pair thresholds. Clearly, if the light stop mass
is smaller, the effects of the $CP$--violating phase $\Phi$ could 
be observed for smaller Higgs--boson masses.

The most crucial observation of the present work is that  
the $CP$--violating neutral Higgs--boson mixing cannot be neglected,
but should be included together with the third--generation sfermion 
loop contributions for reliable estimates of the Higgs--boson 
production rates in the presence of the non--trivial $CP$--violating 
phases. 

We conclude by noting that similar phenomena appear in the
production of the Higgs bosons in two--photon fusion 
\cite{TWOPHOTON,LINEAR} as well as the two--photon decays of the Higgs 
boson. In this case, high energy colliding beams of linearly polarized 
photon beams, which can be generated by Compton back--scattering of 
linearly polarized laser light on electron/positron bunches of $e^+e^-$ 
linear colliders \cite{GKPST},
are expected to provide a clear--cut, direct means for determining the
the $CP$ properties of the produced Higgs bosons. We will discuss
these crucial issues in detail in a future publication \cite{FUTURE}.

\vskip 0.3cm

\section*{Acknowledgments}

The work of SYC was supported by the Korea Science and Engineering 
Foundation (KOSEF) through the KOSEF--DFG large collaboration project, 
Project No.~96--0702--01-01-2.

%

%



\newcommand{\nappend}{\setcounter{equation}{0}
\def\theequation{\rm{A}.\arabic{equation}}\section*}

\nappend{Appendix A: Feynman Rules}

In this Appendix, we list all the Feynman rules needed for the present
work on the Higgs--boson production via gluon--gluon fusion.

Firstly, the interactions of a gluon to fermions are given by
\begin{eqnarray}
{\cal L}_{gff}=-g_s\bar{f}_I\gamma^\mu ({\rm T}^a)_{IJ} f_J G^a_\mu\,,
\end{eqnarray}
and those of gluons to sfermions by
\begin{eqnarray}
{\cal L}_{g\tilde{f}\tilde{f}}=-ig_s({\rm T}^a)_{IJ}
   \tilde{f}^*_{iI}\stackrel{\leftrightarrow}{\partial_\mu}\tilde{f}_{iJ}
   G^{\mu a}
   +g^2_s({\rm T}^a{\rm T}^b)_{IJ}\tilde{f}^*_{iI}\tilde{f}_{iJ}
    G^a_\mu G^{\mu b}\,,
\end{eqnarray}
where ${\rm T}^a=\lambda^a/2$ $(a=1\,{\rm to}\,8)$ are the eight SU(3)
generators in the adjoint representaion, and $G^a_\mu$ the corresponding
eight gluon fields.

Secondly, the interactions of the neutral Higgs--boson fields with fermions 
are described by the Lagrangian
\begin{eqnarray}
{\cal L}_{H\bar{f}f}&=&-\frac{e\, m_f}{2m_W s_WR^f_\beta}\, 
\bar{f}\left[v^i_f-i\bar{R}^f_\beta a^i_f\gamma_5\right]f\, H_i
\equiv  \bar{f}(g^i_{sff}+i\gamma_5 g^i_{pff})f\, H_i\,, \\
R^f_\beta &=&\left\{\begin{array}{c} c_\beta  \\ 
s_\beta\end{array}\right.\,,
\  \
\bar{R}^f_\beta =\left\{\begin{array}{c} s_\beta \\ 
c_\beta\end{array}\right.\,,
\  \
v^i_f=\left\{\begin{array}{c} O_{2,4-i}\\ O_{3,4-i} \end{array}\right.\,,
a^i_f=\left\{\begin{array}{c} O_{1,4-i}\\ O_{1,4-i} \end{array}\right.\,,
\  \
\begin{array}{l} {\rm for}\ \ f=(l:d) \\ {\rm for}\ \ f=(u)  
\end{array}\,.
\end{eqnarray}
Obviously, the Higgs--fermion--fermion couplings are significant for the 
third--generation fermions, $t$, $b$ and $\tau$. We readily see that the 
$CP$--violating neutral Higgs--boson mixing induces a simultaneous coupling of 
$H_i$ ($i=1,2,3$) to $CP$--even and $CP$--odd fermion bilinears $\bar{f}f$
and $\bar{f} i\gamma_5 f$. This can lead to a sizeable phenomena of $CP$
violation in the Higgs decays into polarized top--quark or tau--lepton pairs.

Thirdly, the Feynman rules for the Higgs--sfermion--sfermion vertices,
involving all the mixing and phases and including the phase rotations
of the sfermion fields, can be written in terms of 
$\tilde{C}^f_{\alpha;\beta\gamma}$ as
\begin{eqnarray}
{\cal L}_{H_i\tilde{f}_j\tilde{f}_k}&=&g^i_{\tilde{f}_j\tilde{f}_k}
\tilde{f}^*_j \tilde{f}_k H_i\,,\nonumber\\
g^i_{\tilde{f}_j\tilde{f}_k}&=& C^f_{\alpha;\beta\gamma}
O_{\alpha, 4-i} ( P_f\, U_f)^*_{\beta j} (P_f\,U_f)_{\gamma k}
= \tilde{C}^f_{\alpha;\beta\gamma}
O_{\alpha, 4-i} (U_f)^*_{\beta j} (U_f)_{\gamma k}\,.
\end{eqnarray}
where $\alpha$ denotes three neutral Higgs--boson fields 
$\{a,\phi_1,\phi_2\}$ and $\{\beta,\gamma\}$ denotes the chiralities 
$\{L,R\}$.
The chiral couplings $\tilde{C}^f_{\alpha;\beta\gamma}$ can be obtained in a
rather tedious but straightforward way and, for the scalar--top quarks, 
the couplings are given by
\begin{eqnarray}
&& \tilde{C}^t_{a;LL}=0\,,\nonumber\\
&& \tilde{C}^t_{a;LR}=i\frac{gm_t}{2m_W s_\beta}(c_\beta |A_t|
+ s_\beta|\mu|\,{\rm e}^{i\Phi_{A_t\mu}})\,,
\nonumber\\
&& \tilde{C}^t_{a;RL}=(C^t_{a;LR})^*\,,\nonumber\\
&& \tilde{C}^t_{a;RR}=0\,,\\
&& \tilde{C}^t_{\phi_1;LL}=-\frac{gm_W}{c^2_W} 
c_\beta \left(\frac{1}{2}-\frac{2}{3}s^2_W\right)\,,
\nonumber\\
&& \tilde{C}^t_{\phi_1;LR}=\frac{gm_t}{2m_W s_\beta}|\mu|\, 
{\rm e}^{i\Phi_{A_t\mu}}\,,\nonumber\\
&& \tilde{C}^t_{\phi_1;RL}=(C^t_{\phi_1;LR})^*\,,\nonumber\\
&& \tilde{C}^t_{\phi_1;RR}=-\frac{gm_W}{c^2_W} c_\beta\frac{2}{3}s^2_W\,,\\
&& \tilde{C}^t_{\phi_2;LL}=-\frac{gm^2_t}{m_W s_\beta}
+\frac{gm_W}{c^2_W}s_\beta
\left(\frac{1}{2}-\frac{2}{3}s^2_W\right)\,,\nonumber\\
&& \tilde{C}^t_{\phi_2;LR}=-\frac{gm_t}{2m_W s_\beta} |A_t|\,,\nonumber\\
&& \tilde{C}^t_{\phi_2;RL}=(C^t_{\phi_2;LR})^*\,,\nonumber\\
&& \tilde{C}^t_{\phi_2;RR}=-\frac{gm^2_t}{m_W s_\beta}
+\frac{gm_W}{c^2_W} s_\beta\frac{2}{3}s^2_W\,,
\end{eqnarray}
and, for the scalar--bottom quarks, they are given by
\begin{eqnarray}
&& \tilde{C}^b_{a;LL}=0\,,\nonumber\\
&& \tilde{C}^b_{a;LR}=i\frac{gm_b}{2m_W c_\beta}(s_\beta |A_b|
+ c_\beta|\mu|\,{\rm e}^{i\Phi_{A_b\mu}})\,,
\nonumber\\
&& \tilde{C}^b_{a;RL}=(C^b_{a;LR})^*\,,\nonumber\\
&& \tilde{C}^b_{a;RR}=0\,,\\
&& \tilde{C}^b_{\phi_1;LL}=-\frac{gm^2_b}{m_W c_\beta}
-\frac{gm_W}{c^2_W} 
c_\beta \left(-\frac{1}{2}+\frac{1}{3}s^2_W\right)\,,
\nonumber\\
&& \tilde{C}^b_{\phi_1;LR}=-\frac{gm_b}{2m_W c_\beta} |A_b|\,,\nonumber\\
&& \tilde{C}^b_{\phi_1;RL}=(C^b_{\phi_1;LR})^*\,,\nonumber\\
&& \tilde{C}^b_{\phi_1;RR}=-\frac{gm^2_b}{m_W c_\beta}
+\frac{gm_W}{c^2_W} c_\beta\frac{1}{3}s^2_W\,,\\
&& \tilde{C}^b_{\phi_2;LL}=\frac{gm_W}{c^2_W}s_\beta
\left(-\frac{1}{2}+\frac{1}{3}s^2_W\right)\,,\nonumber\\
&& \tilde{C}^b_{\phi_2;LR}=-\frac{gm_b}{2m_W c_\beta}|\mu|\, 
{\rm e}^{i\Phi_{A_b\mu}}\,,
\nonumber\\
&& \tilde{C}^b_{\phi_2;RL}=(C^b_{\phi_2;LR})^*\,,\nonumber\\
&& \tilde{C}^b_{\phi_2;RR}= -\frac{gm_W}{c^2_W} s_\beta\frac{1}{3}s^2_W\,,
\end{eqnarray}
with $\Phi_{A_t\mu}=\Phi_{A_t}+\Phi_\mu+\xi$ and $\Phi_{A_b\mu}=\Phi_{A_b}
+\Phi_\mu+\xi$.
Therefore, if the univesal phase $\Phi_{A_t}=\Phi_{A_b}$ is taken, the
induced phase $\xi$ depends only on the combination $\Phi_A+\Phi_\mu$ and
as a result all the Higgs--sfermion--sfermion couplings are determined 
by only one $CP$--odd phase $\Phi=\Phi_A+\Phi_\mu$, the sum of the two 
phases $\Phi_A$ and $\Phi_\mu$.

\mbox{ }

%
\begin{figure}
\begin{center}
\hbox to\textwidth{\hss\epsfig{file=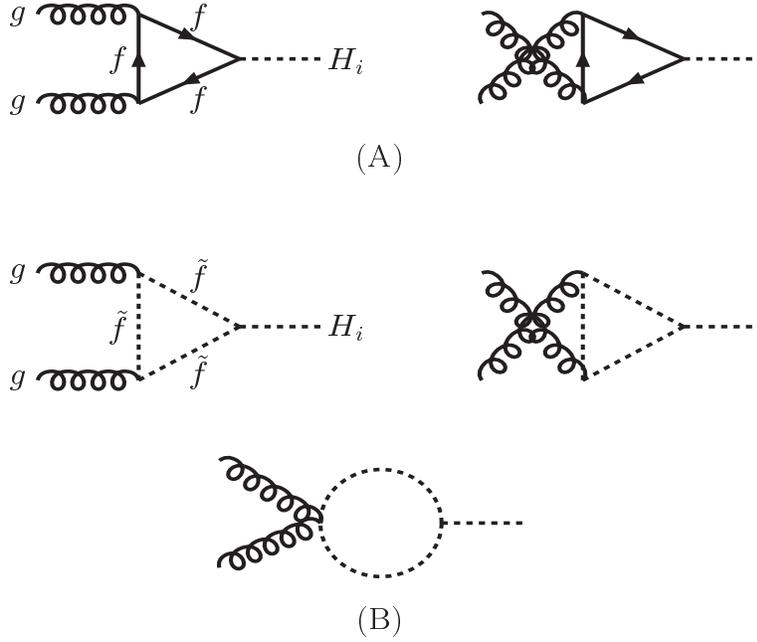,width=16cm,height=20cm}\hss}
\end{center}
  \vskip -6.5cm
\caption{The one--loop diagrams contributing to the process $gg\rightarrow
         H_i$; (A) SM--like contributions from fermions $f$, in particular,
	 top and bottom quarks, and
	 (B) SUSY--like contributions from sfermions $\tilde{f}$, 
	 in particular,
	 scalar top and bottom quarks
	 in the MSSM.}
\label{fig:fig1}
\end{figure}

\newpage
\mbox{ }

%
\begin{figure}
\begin{center}
\hbox to\textwidth{\hss\epsfig{file=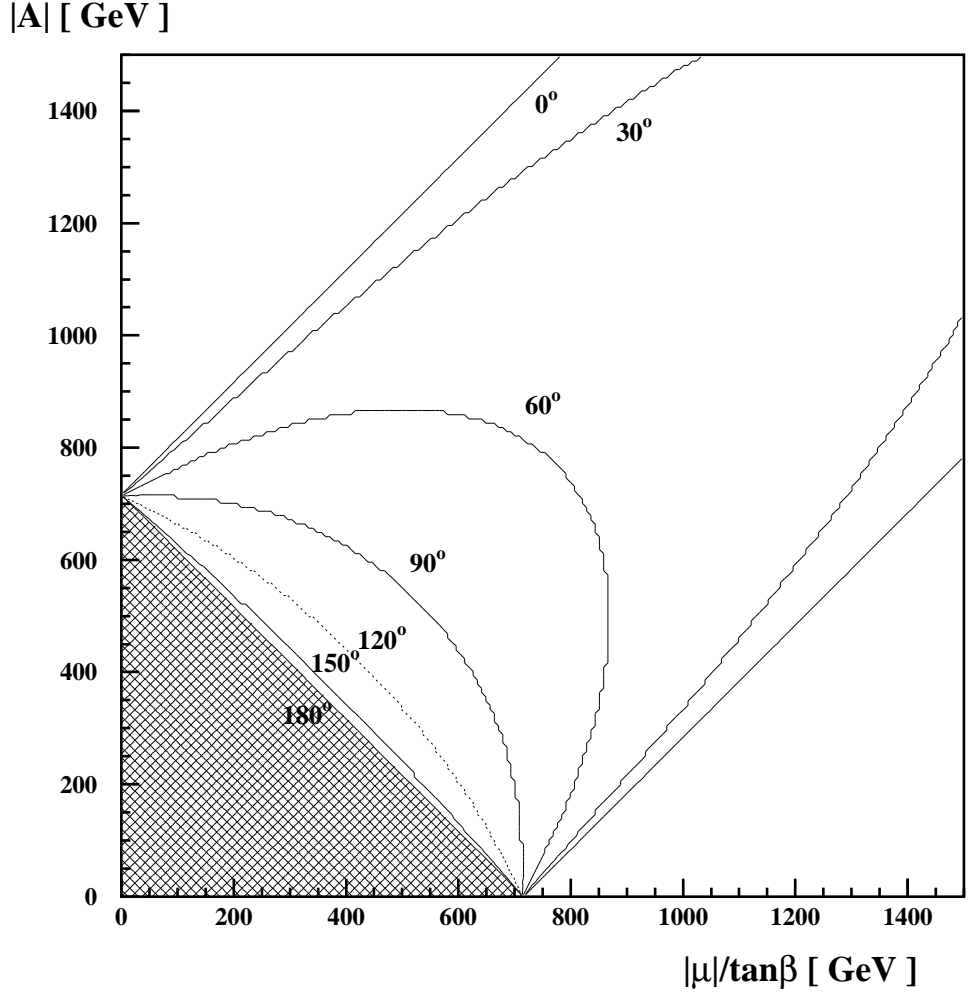,width=16cm,height=16cm}\hss}
\end{center}
\caption{The allowed regime for the CP--violating phase $\Phi_{A\mu}= 
         \Phi_A+\Phi_\mu+\xi$ on the parameter space of
	 two real SUSY parameters $|A|$ and $|\mu|/\tan\beta$ for a
	 fixed value of $\tan\beta=3$. The lower-left hatched triangular 
	 area allows the full range
	 for the phase $\Phi_{A\mu}$, while the lower-right and upper-left
	 regime are not allowed due to the obvious condition 
	 $|\cos\Phi_{A\mu}|\leq 1$. The number of each contour line 
	 denotes the maximally
	 allowed value of the phase for given $|A|$ and $|\mu|/\tan\beta$.}
\label{fig:fig2}
\end{figure}

\newpage
\mbox{ }

%
\begin{figure}
\begin{center}
\hbox to\textwidth{\hss\epsfig{file=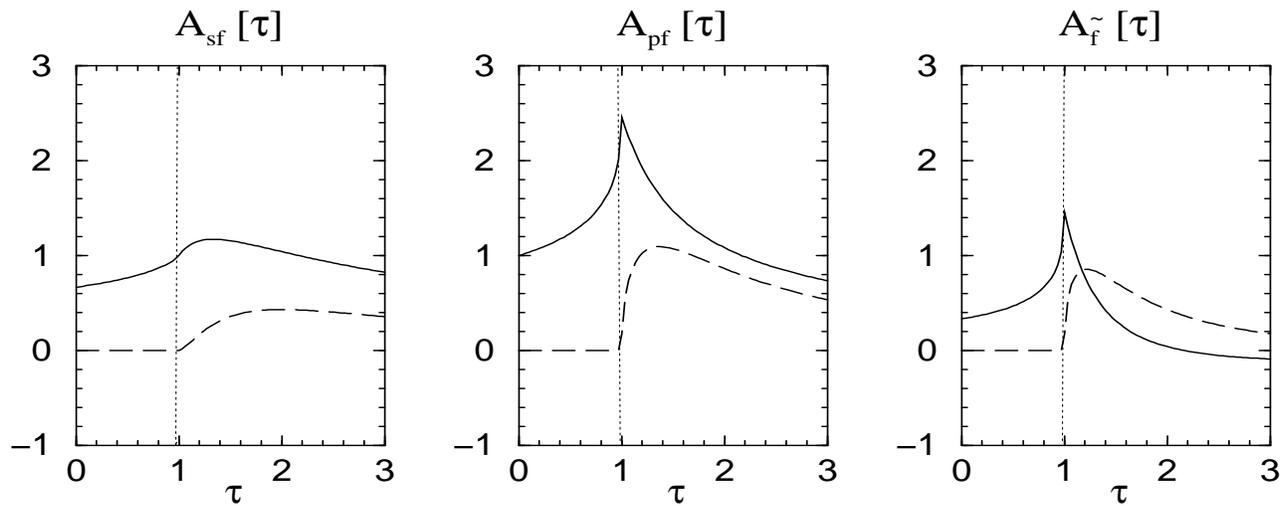,width=17cm,height=7cm}\hss}
\end{center}
\caption{The real (solid) and imaginary (dashed) parts of the form factors, 
         $A_{sf}$ (left), $A_{pf}$ (middle), and $A_{\tilde{f}}$ (right), 
	 with respect to the Drell--Yan variable $\tau$. The vertical 
	 dotted line is to denote the thresholds for Higgs-boson decays 
	 into a fermion--pair or a sfermion--pair. The real parts have their
	 peaks at $\tau=1$ and the imaginary parts are non--zero only
	 above the threshold.}
\label{fig:fig3}
\end{figure}

\newpage
\mbox{ }

%
\begin{figure}
\begin{center}
\hbox to\textwidth{\hss\epsfig{file=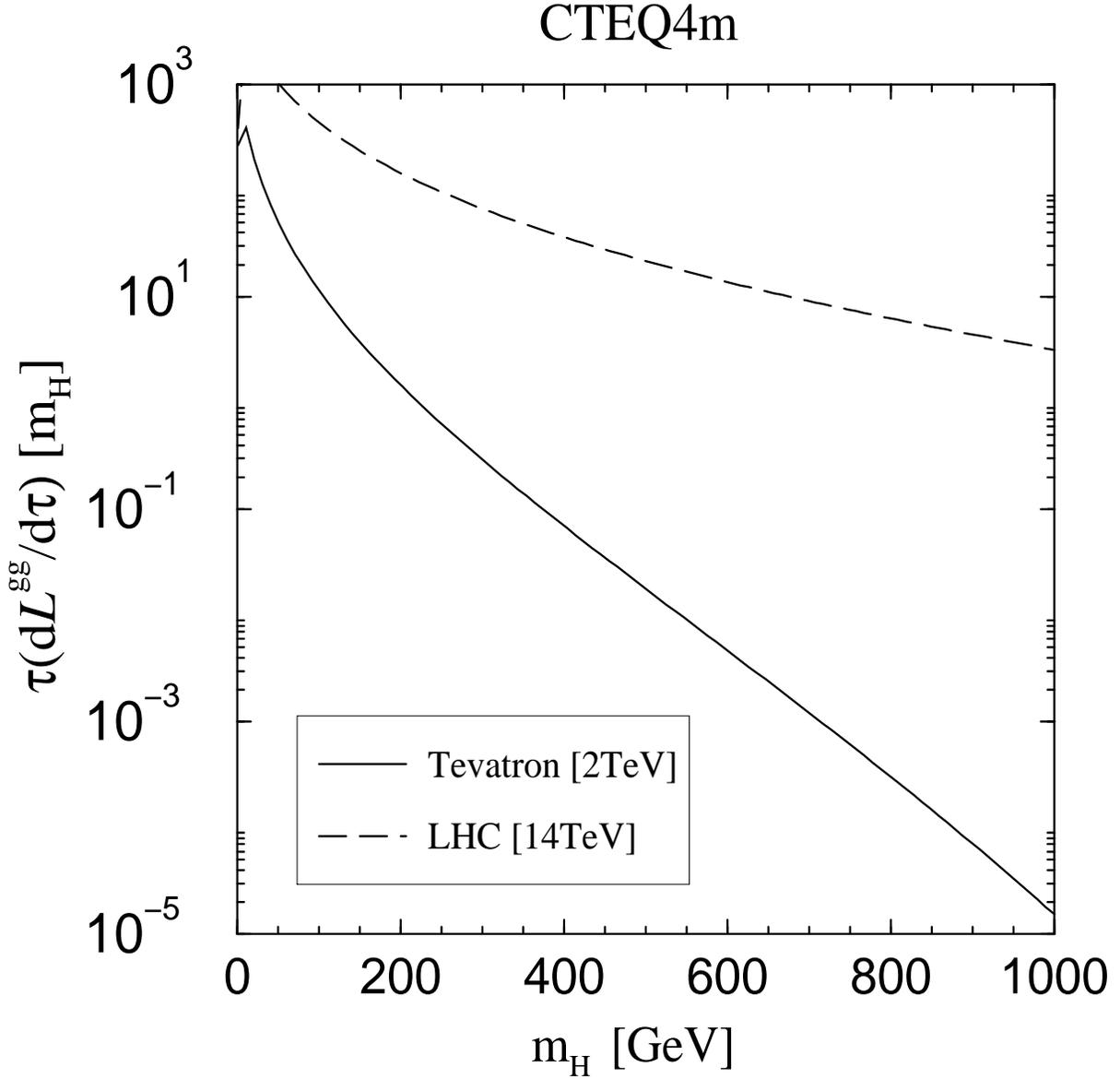,width=16cm,height=16cm}\hss}
\end{center}
\caption{The gluon--gluon luminosity function with respect to the 
         Higgs--boson mass $m_H$ at the Tevatron (solid) with $\sqrt{s}=2$ TeV
	 and at the LHC (dashed) with $\sqrt{s}=14$ TeV. The Drell-Yan 
	 variable $\tau$ denotes $m^2_H/s$ by using the CTEQ4m 
	 parameterization for gluon densities. The luminosity function is
	 larger at the LHC by two orders of magnitude than at the Tevtaron.
	 It decreases very abruptly with increasing Higgs--boson mass
	 $m_H$ at the Tevatron, while it does not decrease so much at the
	 LHC.} 
\label{fig:fig4}
\end{figure}

\newpage
\mbox{ }

%
\begin{figure}
\begin{center}
\hbox to\textwidth{\hss\epsfig{file=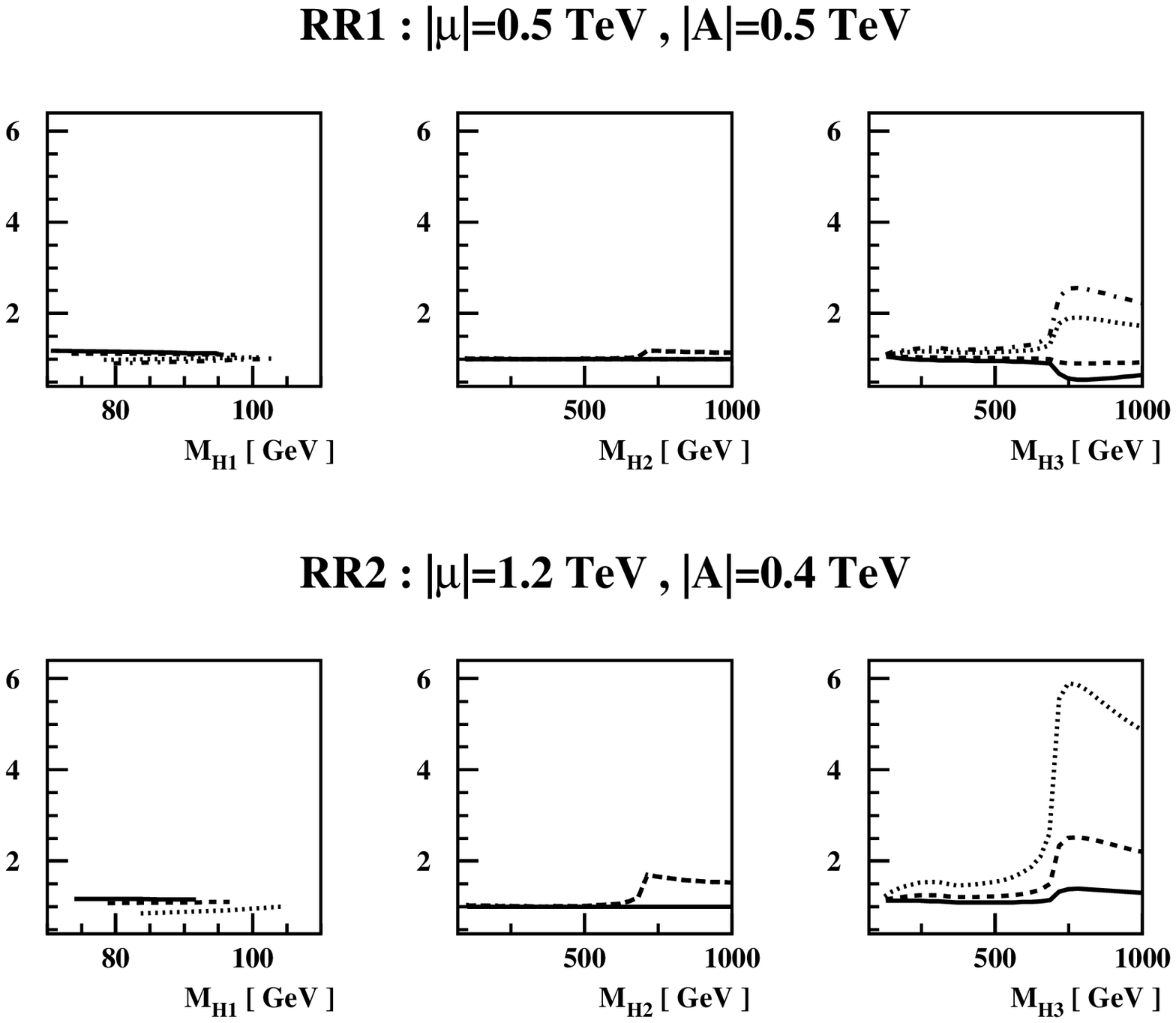,width=16cm,height=18cm}\hss}
\end{center}
  \vskip -1.cm
\caption{ $\hat{\sigma}^0_{LO}[f+\tilde{f}]/\hat{\sigma}^0_{LO}[f]$
         with respect to the mass 
	 of the produced Higgs boson.
         The upper three frames are in the 
	 scenario $RR1$, while the lower three frames are in the 
	 scenario $RR2$. 
	 The phase $\Phi=\Phi_A+\Phi_\mu$ is taken to be $0^0$ (solid),
	 $60^0$ (dashed), $120^0$ (dotted), and $180^0$ (dot--dashed).
	 Note that in $RR2$ the phase angle only up to about $140^0$ is 
	 not allowed.}
\label{fig:fig5}
\end{figure}

\newpage
\mbox{ }

%
\begin{figure}
\begin{center}
\hbox to\textwidth{\hss\epsfig{file=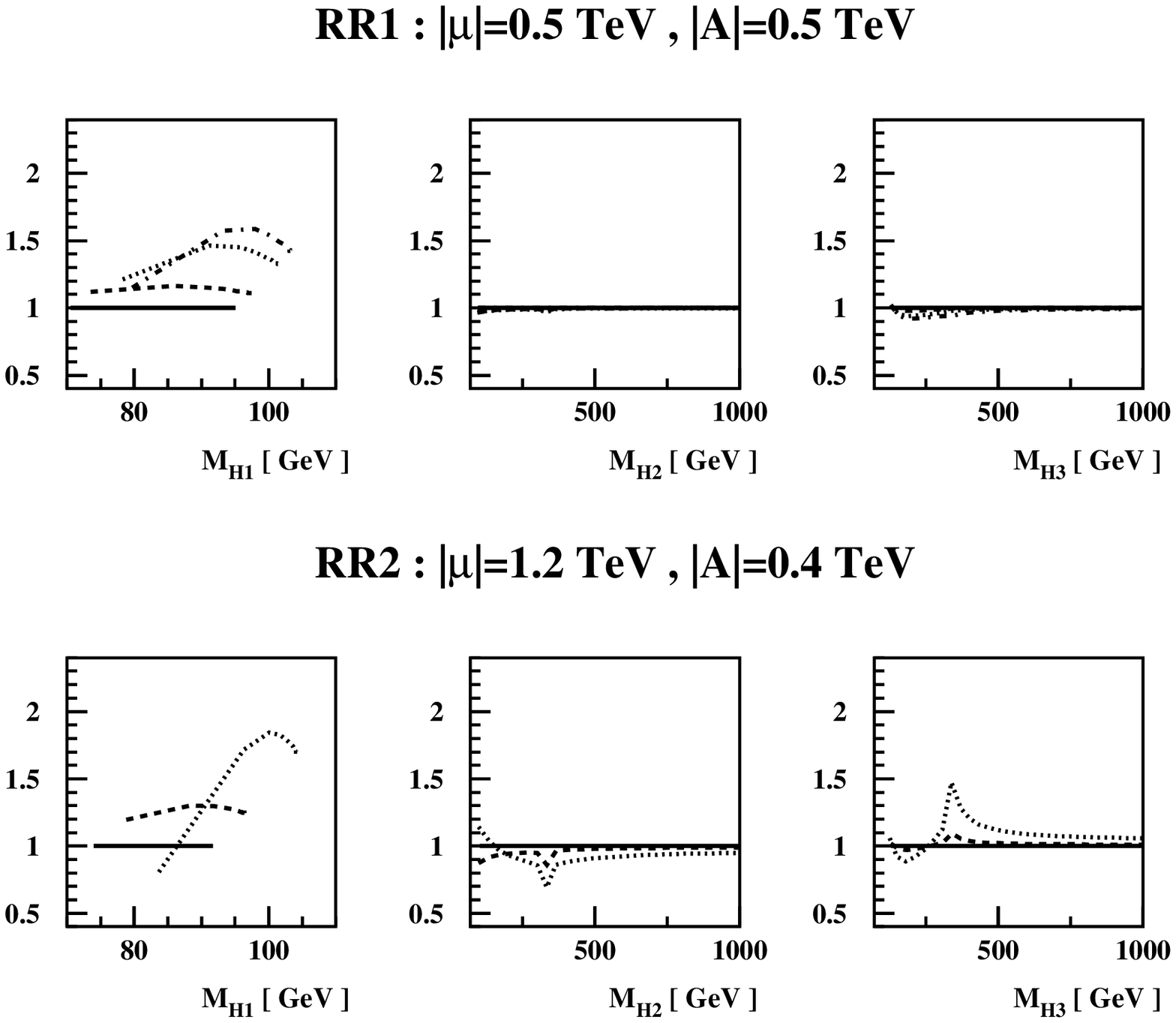,width=16cm,height=18cm}\hss}
\end{center}
  \vskip -1.cm
\caption{$\hat{\sigma}_{LO}[f]/\hat{\sigma}^0_{LO}[f]$
	 with respect to the mass of the produced Higgs boson.
	 The upper three frames are in the scenario $RR1$, 
	 while the lower three frames are in the scenario $RR2$. 
	 The phase $\Phi=\Phi_A+\Phi_\mu$ is taken to be $0^0$ (solid),
	 $60^0$ (dashed), $120^0$ (dotted), and $180^0$ (dot--dashed).}
\label{fig:fig6}
\end{figure}

\newpage
\mbox{ }

%
\begin{figure}
\begin{center}
\hbox to\textwidth{\hss\epsfig{file=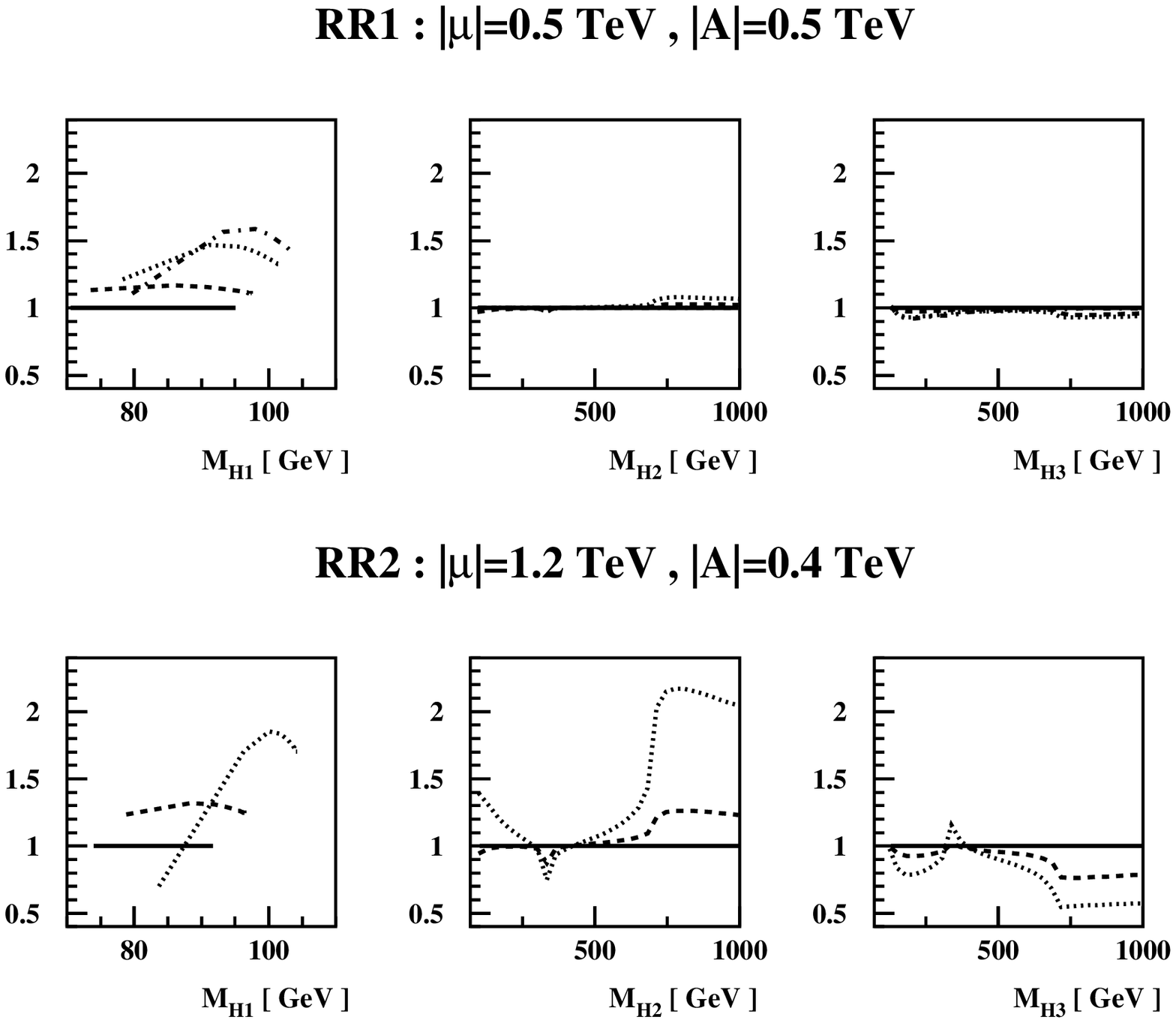,width=16cm,height=18cm}\hss}
\end{center}
  \vskip -1.cm
\caption{$\hat{\sigma}_{LO}[f+\tilde{f}]/\hat{\sigma}^0_{LO}[f
         +\tilde{f}]$
	 with respect to the mass of the produced Higgs boson. 
	 The upper three frames are in the scenarios $RR1$, 
	 while the lower three frames are in the scenario $RR2$. 
	 The phase $\Phi=\Phi_A+\Phi_\mu$ is taken to be $0^0$ (solid),
	 $60^0$ (dashed), $120^0$ (dotted), and $180^0$ (dot--dashed).}
\label{fig:fig7}
\end{figure}

\newpage
\mbox{ }

%
\begin{figure}
\begin{center}
\hbox to\textwidth{\hss\epsfig{file=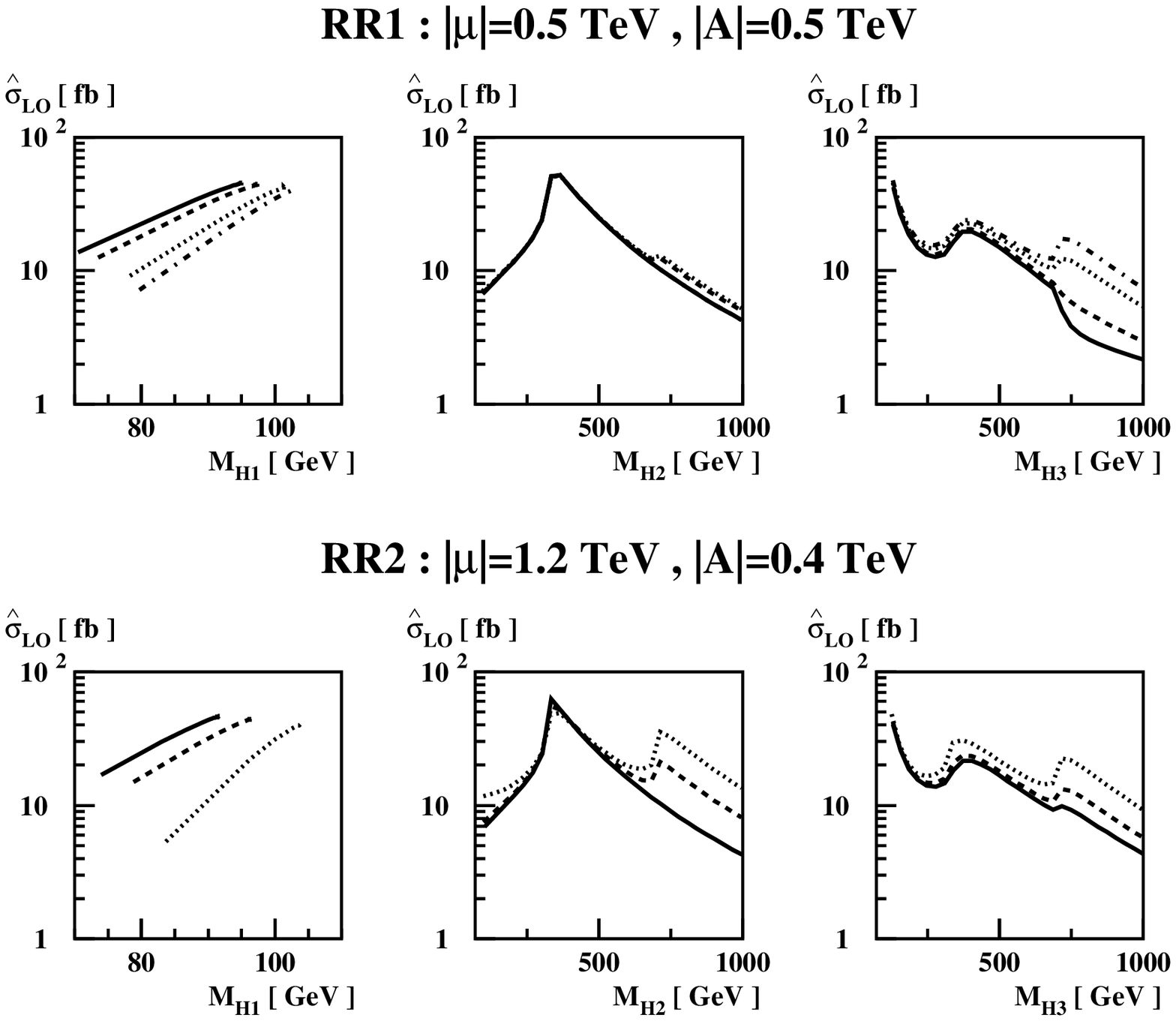,width=16cm,height=18cm}\hss}
\end{center}
  \vskip -1.cm
\caption{$\sigma_{LO}(gg\rightarrow H_i)$ ($i=1,2,3$) with the neutral 
	 Higgs--boson mixing as well as the
	 stop/sbottom loop corrections with respect to the mass of 
	 the produced Higgs boson. 
	 The upper three frames are in the scenarios $RR1$, 
	 while the lower three frames are in the scenario $RR2$. 
	 The phase $\Phi=\Phi_A+\Phi_\mu$ is taken to be $0^0$ (solid),
	 $60^0$ (dashed), $120^0$ (dotted), and $180^0$ (dot--dashed).}
\label{fig:fig8}
\end{figure}
\vfill\eject

\end{document}